\documentclass[twocolumn,showpacs,preprintnumbers,amsmath,amssymb]{revtex4}
\usepackage{graphicx}% Include figure files
\input epsf
\usepackage{dcolumn}% Align table columns on decimal point
\usepackage{bm}% bold math
\usepackage[pdftex,
            pdfauthor={A.V. Yurov, V.A. Yurov},
            pdftitle={On the question of generalized Darboux transformations and QNM spectra in  Schwarzschild  black hole perturbation theory},
            pdfsubject={},
            pdfkeywords={Darboux transformation, Schwarzschild black hole},
            pdfproducer={Latex with hyperref},
            pdfcreator={pdflatex}]{hyperref}
\hypersetup{colorlinks=true, citecolor=blue, linkcolor=red}
\newcommand{\beq}{\begin{equation}}
\newcommand{\beqn}{\begin{equation*}}
\newcommand{\enq}{\end{equation}}
\newcommand{\enqn}{\end{equation*}}

\newcommand{\R}{{\mathbb R}}

\renewcommand{\Im}{\text{\rm Im}}
\renewcommand{\l}{\lambda}

\begin{document}

%\preprint{APS/123-QED}

\title{A look at the generalized Darboux transformations for the quasinormal spectra in Schwarzschild black hole perturbation theory: just how general should it be?}

\author{A.V. Yurov}
\email{AIurov@kantiana.ru}
%This line break forced with \textbackslash\textbackslash%
\author{V.A. Yurov}%
 \email{vayt37@gmail.com}
\affiliation{%
Immanuel Kant Baltic Federal University, Institute of Physics, Mathematics and Informational Technology,
 Al.Nevsky St. 14, Kaliningrad, 236041, Russia
\\
}%
%\altaffiliation[Also at ]{The Theoretical Physics Department, Kaliningrad State University, A. Nevskogo str., 14, 236041,  Russia.}

%\affiliation{

\date{\today}% It is always \today, today,

\begin{abstract}
In this article we take a close look at three types of transformations usable in the Schwarzschild black hole perturbation theory: a standard (DT), a binary (BDT) and a generalized (GDT) Darboux transformations. In particular, we discuss the absolutely crucial property of {\em isospectrality} of the aforementioned transformations which guarantees that the quasinormal mode (QNM) spectra of potentials, related via the transformation, completely coincide. We demonstrate that, while the first two types of the Darboux transformations (DT and BDT) are indeed isospectral, the situation is wildly different for the GDT: it violates the isospectrality requirement and is therefore only valid for the solutions with just one fixed frequency. Furthermore, it is shown that although in this case the GDT does provide a relationship between two arbitrary potentials (a short-ranged and a long-ranged potentials relationship being just a trivial example), this relationship ends up being completely formal.  Finally, we consider frequency-dependent potentials. A new generalization of the Darboux transformation is constructed for them and it is proven (on a concrete example) that such transformations are also not isospectral. In short, we demonstrate how a little, almost incorporeal flaw may become a major problem for an otherwise perfectly admirable goal of mathematical generalization.
\end{abstract}

%Use showkeys class option if keyword
\maketitle

\section{\label{sec:level1}Introduction
%First-level heading:\protect\\ The line
%break was forced \lowercase{via} \textbackslash\textbackslash
}
The black holes are truly fascinating objects. Ever since their premature birth in 1915 (more than 50 years prior to John Wheeler coining the very name \footnote{Reportedly after  a suggestion made by one of his students.}), when Karl Schwarzschild has managed to find an exact solution to the Esintein's equations describing the space-time continuum surrounding a point mass \cite{Schwarzschild}, it has never ceased to astonish and amaze the scientists who study them. One of its many peculiarities is that it is a sort of an object which, being invisible by itself, can nevertheless be detected by the influence it exerts upon the intermediary observable agent. The agents can be of different sorts: the neighbouring stars in a binary system \cite{SLD}; the stars near the galactic nucleus orbiting the supermassive black hole in the center of the galaxy \cite{GET}; the radiation emitted by the accretion discs formed around the black holes \cite{Munoz}; or just the photons travelling from distant stars, whose trajectories are distorted in a gravitational lens formed by a curved space-time around the black hole \cite{Bozza}. What unites all these phenomena is the fact that to be observable they all must occur in a sufficiently close proximity of an event horizon, inevitably {\em perturbing} it. Therefore, we end up with a curious maxima: for a black hole to be visible it must be perturbed; when it is not perturbed it is invisible. This naturally implies that all physically meaningful (i.e. observable) black holes must be studied in their perturbed form \footnote{Even the black holes with literally {\em nothing} around them are perturbed by the very vacuum they are embedded in, producing a famous Hawking radiation.}, in which the black hole emits the gravitational waves, dominated by the {\em quasi-normal modes} (QNM): the proper oscillations with a dampened amplitude (for some very good in-depth reviews on the subject see \cite{Kokkotas}, \cite{Nollert}, \cite{Berti} and \cite{Konoplya}).

Last year a team of theoretical physicists, Kostas Glampedakis, Aaron D. Johnson and  Daniel Kennefick (the team we will from now on refer to as GJK) has received and published \cite{GJK-2017} one truly remarkable result: they have managed to demonstrate that the solutions of the Zerilli \cite{Z-1970} and Regge-Wheeler \cite{RW-1957} equations, that describe different gravitational perturbations of the Schwarzschild geometry, are actually related to each another by means of the Darboux transformations (DT).  This extremely interesting fact proves, among other things, that the QNM spectra of these seemingly extremely different problems coincide. We note that for standard quantum mechanical problems the isospectrality is usually preserved by a special choice of a support function  (in  \cite{GJK-2017} it is denoted by $X_*$), which is chosen to be strictly positive for the entire domain: $-\infty<x<\infty$.  However, the isospectral property of DT for those eigenvalues from a QNM spectra derived for the theory of black holes is somewhat trickier, and requires a different approach. The difficulty here is that while the DT formally allows one to generate some new solutions for a new equation with a pre-chosen value of the spectral parameter, it is nevertheless not necessarily true that these solutions (the old and the new ones) will have similar spectra. In quantum mechanics, this hurdle can be eliminated by a simple additional assumption that the new solutions must belong to $L^2(-\infty,\infty)$, whereas  in the black hole perturbation theory case one shall instead impose the condition that $A_{in}(\omega)=0$, where  $A_{in}(\omega)$ is  the coefficient of the exponential factor  $\exp(-i\omega x)$ when  $r\to\infty$ ($x$  is the standard tortoise coordinate: $x=r+2M\log|r-2M|$ for the Schwarzschild  black hole).

Without dwelling too much on what has been achieved in \cite{GJK-2017}, the authors there attempted to establish a connection with a third equation describing the Schwarzschild perturbations -- the Bardeen-Press equation \cite{BP-1973} (this equation may be obtained from the Teukolsky equation for the Kerr perturbation in the non-rotating limit). But while pursuing this goal, the authors have stumbled upon the apparent roadblock: the Zerilli and Regge-Wheeler potentials are both short-ranged ($V_{_{RW}}\sim V_{_Z}\sim 1/r^2$ at $r\to\infty$), while the Bardeen-Press potential is actually long-ranged: $V_{_{BP}}\sim 1/r$  at $r\to\infty$. To make matters even worse, the  Bardeen-Press potential  is also complex-valued and frequency-dependent.   In other words, the  Bardeen-Press equation has the form
(throughout the article, a prime denotes differentiation with respect to the  tortoise coordinate $ x $):
\begin{equation}
\Phi''(x;\omega)+\left(\omega^2
-V_{_{BP}}(r;\omega)\right)=0,
\label{BP-eq}
\end{equation}
where the explicit form of the potential $V_{_{BP}}$ will not be important for the purposes of this article.  Using the beautiful integro-differential relation between potentials coupled by DT (i.e. applying the B\"acklund transformations for the  KdV hierarchy \cite{Matveev-Salle}), GJK have presented an elegant proof of the fact that it is impossible to obtain a long-ranged potential from a short-ranged  one via the DT. But then, in a desperate attempt to circumvent this problem, GJK have tried to introduce a so called generalized Darboux transformations (GDT). They alluded to the fact that in the original work of Darboux in 1882, these transformations were introduced precisely in such a generalized form \cite{Darboux-1882}~\footnote{Our sincerest appreciation goes to a nameless hero who has submitted the electronic version of this work to the arXiv!}. GDT contains an additional function which was supposed to bypass the restrictions imposed by DT and to succeed in connecting the short-ranged   and   long-ranged  potentials via the GDT. Unfortunately, that's where the problems starts to crop up. In fact, after all the praise wholeheartedly given to the first half of the article \cite {GJK-2017} (as the authors have indeed managed to solve the very infamous puzzle that surprised Chandrasekhar so much), sadly, we are forced to strongly criticize the entire idea of GDT's effectiveness in comparison with the standard DT. \cite {GJK-2017}.

We show that these transformations do not meet the two most important conditions that distinguish DT precisely among all the transformations generated by first-order differential operators and therefore
do not correspond to the isospectral property. Therefore, they only associate solutions with one given frequency QNM spectra. Moreover, we demonstrate that when we restrict our choice of transformations to those that connect the functions having identical frequencies, then any need in GDT simply disappears: ordinary DTs allow one to associate solutions for two arbitrary  potentials (the short-ranged and long-ranged potentials relationship is just a trivial example). However, from a practical point of view, this fact means nothing -- even knowing the QNM spectrum of the original problem, we can not say anything about the spectrum of the second. In other words, unlike classical DTs, GDTs are purely formal and, with a few notable exceptions (see Sec. \ref{GDT}) their effectiveness leaves much to be desired. Among those exceptions one might name, for example, the Chandrasekhar-Sasaki-Nakamura transformation (essentially a GDT) which has been developed by Sasaki and Nakamura \cite{Sasaki} to calculate the gravitational radiation emitted to infinity by producing a short-ranged potential known as the Sasaki-Nakamura potential (see Section 2.2 of \cite{Sasaki_Tagoshi}). Another example is interesting relationship between the raising and lowering operators for spin-weighted spheroidal harmonics, which also has a form of GDT \cite{Shah_Whiting}. However, what we are interested in for this article is somewhat different from both of those cases, as we primarily concentrate on the intrinsic properties of the QNM when the potentials are produced by DT and GDT. As we shall see, for this particular task the relative effectiveness of GDT is rather lacking (although there might still be some exceptions, as will be discussed below).

The paper is organized in the following fashion. We will begin the next section by describing the two main properties of DT that distinguish it from the more general transformations, particularly by preserving the isospectrality. Then we complete the proof for the isospectrality of the DT binding potentials $V _ {_ {RW}} $ and $ V _ {_ Z} $, started in \cite {GJK-2017}. The description and critics of the GDT are provided in Sec. \ref{sec:GDT}. In Sec. \ref{sec:Frequency} we give general formulas for the Darboux-like transformations in the case of frequency-dependent potentials and show  that these transformations are not isospectral in general. We discuss our findings and provide some final remarks in the Conclusion.

\section{Darboux transformations} \label{sec:DT}

In this paper we use the notation traditional for  DT's papers. For convenience, we give a brief summary of the notations used in \cite {GJK-2017} and here:  $y(x), X(x)$ we denote by $\psi(x;\lambda)$, $u=X_*\to \phi(x;\mu)$, $\omega^2 \to \lambda$, $\omega_*^2\to \mu$, $Z(x)\to\psi^{(1)}(x;\lambda;\mu)$,$V_{_{RW}}\to V$, $V_{_Z}\to V^{(1)}$, $\beta(x)\to A(x)$, $q(x)=\lambda-V$, $Q(x)=V^{(1)}-\lambda$, $f\to -\sigma$.

DT for linear differential equations of the second order are well known. We are primarily interested in the equations of the Schr\"odinger type:
\begin{equation}
\psi''(x;\lambda)+\left(\lambda-V(x)\right)\psi(x;\lambda)=0.
\label{Scr}
\end{equation}
Sometimes (in situations where this does not cause confusion) we will not specify independent variables, for example  $\psi(x;\lambda)=\psi$  and so on. Let  $\phi=\phi(x;\mu)$ (in what follows -- {\bf the support function}) be a solution of (\ref{Scr}) for the same potential but a different value of the spectral parameter:
\begin{equation}
\phi''(x;\mu)+\left(\mu-V(x)\right)\phi(x;\mu)=0.
\label{Scr-1}
\end{equation}
Here it is extremely important that the potential does not depend on the spectral parameter (it can depend on $\mu$, though). It is convenient to introduce the function $\sigma=\left(\log\phi\right)'$, which is the solution of the Riccati equation:
\begin{equation}
\sigma'+\sigma^2+\mu-V=0,
\label{Ric}
\end{equation}
to define the DT:
\begin{equation}
\begin{array}{l}
\psi(x;\lambda)\to\psi^{(1)}(x;\lambda;\mu)=\psi'-\sigma\psi,\\
\\
V(x)\to V^{(1)}(x;\mu)=V-2\sigma',\\
\\
\lambda\to \lambda^{(1)}=\lambda.
\end{array}
\label{DT}
\end{equation}
It is easy to show (for example, via the method of factorization), that DT maps $L^2$ into itself (except perhaps for a zero mode corresponding to the spectral parameter $\mu $) if and only if the support function $ \phi$  is positive  everywhere.  In addition, DT is invertible on the subspace $L^2\backslash\ker(DT)$ (we will discuss the significance of a kernel of DT below), so the discrete spectra (bound states in quantum mechanics) of the potentials $ V (x) $ and $ V^ {(1)} (x; \mu) $ are completely identical to each other with the sole exception of a zero mode.

Some authors call the "Darboux transformation"  a more general substitution, of the form
\begin{equation}
\psi\to A(x)\psi'-B(x)\psi.
\label{GDT}
\end{equation}
In  \cite{GJK-2017}, such a transformation with  $A(x)\ne {\rm const}$   is defined as a generalized Darboux transformation (GDT). We will discuss these transformations in detail in the next section,  but now we note two main properties that actually distinguish the Darboux transformations from the general transformations  class of type (\ref{GDT}):

{\bf Property 1}. DT is generated by a linear differential operator of the first order that has a nonzero kernel (zero mode) in the solutions space of the initial equation.

In other words, there must exist a solution of the original equation (the support function $\phi$ in our case) which upon substitution in (\ref {GDT}) yields nothing but zero. It is this property that allows one to derive the  Crum's Wronskian formulas \cite{Crum} and the isospectrality of the DT (including the behavior of the reflection-transmission coefficients, which after DT are simply multiplied by a phase factor).
The origin of this property is easier to understand from the factorizability of the Hamiltonian into a product of two Hermitian conjugate operators $\bf q$ and $\bf q^+$, of whom the rightmost (in the order of multiplication) operator $ {\bf q} = d/dx- \sigma $  generates the Darboux transformation, whereas  the Hermitian conjugate ${\bf q}^+$ defines an inverse transformation (see also \cite{Infeld-Hall}). The  Property 1 is an obvious consequence of the fact that the equation $ {\bf q} \phi = 0 $ always has nontrivial solutions.

{\bf Property 2}.   After the DT new potential  depends only on the value of the spectral parameter of the null mode: $V^{(1)}=V^{(1)}(x;\mu)$.

This property is quite obvious. Indeed, if after the DT (\ref{DT}) $ V^{(1)} = V^{(1)}(x; \mu; \lambda) $, then different values of the $\lambda $  will correspond to {\em different} potentials. Property 2 immediately implies that if the potential explicitly depends on a spectral parameter (e.g., the Bardeen-Press potential  $V_{_{BP}}$  which arises from the Teukolsky equation written for the Kerr perturbations \cite {Teu}), then no transformation of the form (\ref{GDT}) applied to it will have the property of isospectrality. This will be the case of general position even if we call it a Darboux transformations. We will explicitly show this in the fourth section.

Thus, we postpone the critical part of our discussion until the next section, and proceed with our consideration of the isospectrality of Darboux transformation for QNM spectra of the potential  $V_{_{RW}}$ and $V_{_Z}$.  A quasinormal mode (QNM) $\{\omega_n\}$ spectrum is defined in the following way. First, we impose a ${\rm e}^{-i\omega t}$ time-dependence upon the perturbation and demand for it to be vanishing for large $t$. This, of  course, implies that
\begin{equation}
\Im ~\omega_n<0,
\label{Im}
\end{equation}
for all integers $n$. Next, we have to solve the equation
\begin{equation}
\psi''+(\omega^2-V_{_{RW/Z}})\psi=0,
\label{QNM-eq}
\end{equation}
and the solutions shall satisfy the following asymptotic behavior:
\begin{equation}
\begin{array}{l}
\psi(x\to+\infty)=A_{in}(\omega){\rm e}^{-i\omega x}+A_{out}(\omega){\rm e}^{i\omega x},\\
\\
\psi(x\to-\infty)=B_{in}(\omega){\rm e}^{-i\omega x}.
\end{array}
\label{as}
\end{equation}
Thus the solutions of the (\ref{QNM-eq}) has the form
\begin{equation}
\begin{array}{l}
\psi(x>0;\omega)\equiv \psi_+=y_-(x;\omega){\rm e}^{-i\omega x}+y_+(x;\omega){\rm e}^{i\omega x},\\
\\
\psi(x<0;\omega)\equiv \psi_-=z(x;\omega){\rm e}^{-i\omega x},
\end{array}
\label{9-10}
\end{equation}
with the continuity conditions:
\begin{equation}
\begin{array}{l}
z'(0;\omega)-y'_-(0;\omega)-y'_+(0;\omega)=2i\omega y_+(0;\omega),\\
\\
y_-(0;\omega)+y_+(0;\omega)=z(0;\omega).
\end{array}
\label{cc}
\end{equation}
Then
\begin{equation}
\begin{array}{l}
\displaystyle{
A_{in}(\omega)=\lim_{x\to +\infty}y_-(x;\omega)},\\
\\
\displaystyle{
 A_{out}(\omega)=\lim_{x\to +\infty}y_+(x;\omega)},\\
\\
\displaystyle{
 B_{in}(\omega)=\lim_{x\to -\infty}z(x;\omega)},
\end{array}
\label{10-9}
\end{equation}
and QNM spectra $\{\omega_n\}$ is defined as the solution of the equation
\begin{equation}
A_{in}(\omega)=0,
\label{QNM-sp}
\end{equation}
with  the additional condition (\ref{Im}).

For the  Regge-Wheeler potential
\begin{equation}
V=\left(1-\frac{2M}{r}\right)\left(\frac{l(l+1)}{r^2}-\frac{6M}{r^3}\right),
\label{R-W}
\end{equation}
the support function has the form
\begin{equation}
\begin{array}{l}
\displaystyle{
\phi=\left(n+\frac{3M}{r}\right){\rm e}^{-i\omega_*x},\qquad n=\frac{(l-1)(l+2)}{2}},\\
\\
\displaystyle{\omega_*=\sqrt{\mu}=-\frac{in(n+1)}{3M}}.
\end{array}
\label{prop}
\end{equation}
Since the conditon (\ref{Im}) is valid one concludes that $\omega_*\in\{\omega_n\}$.  When $l=0$ or $l=1$ we end up with $\mu=0$, while for $l>1$ the potential vanishes on the horizon, then increases to reach a maximum at
$$
r=\frac{\left(3(l^2+l+3)+\sqrt{9l^4+18l^3-33 l^2-42l+81}\right)M}{2l(l+1)},
$$
after which it decreases asymptotically as $ l(l + 1) / r ^ 2 $, remaining positive everywhere. Using (\ref{prop}) we find  $\sigma $:
$$
\displaystyle{
\sigma=-\frac{3M(r-2M)}{(nr+3M)r^2}-\frac{n(n+1)}{3M}},
$$
 wherein $\sigma(x\to\pm\infty;\mu)=-i\omega_*$.  Substituting  $\sigma$ and (\ref{9-10}) into the  (\ref{DT}) one gets:
\begin{equation}
\begin{array}{l}
\psi_+^{(1)}(x,\omega;\omega_*)=y^{(1)}_{-}(x;\omega;\omega_*){\rm e}^{-i\omega x}+y^{(1)}_+(x;\omega;\omega_*){\rm e}^{i\omega x},\\
\\
\psi_-^{(1)}(x,\omega;\omega_*)=z^{(1)}(x;\omega;\omega_*){\rm e}^{-i\omega x},
\end{array}
\label{gen-1}
\end{equation}
where
\begin{equation}
\begin{array}{l}
y^{(1)}_{\pm}(x;\omega;\omega_*)=y'_{\pm}(x;\omega)-\left(\sigma(x;\omega_*)\mp i\omega\right)y_{\pm}(x;\omega),\\
\\
 y^{(1)}(x;\omega;\omega_*)=z'(x;\omega)-\left(\sigma(x;\omega_*)+ i\omega\right)z(x;\omega).
\end{array}
\label{as-2}
\end{equation}
Thus
\begin{equation}
\begin{split} \label{izo}
A^{(1)}_{in}(\omega;\omega_*) &=i(\omega_*-\omega)A_{in}(\omega),\\
A^{(1)}_{out}(\omega;\omega_*) &=i(\omega_*+\omega)A_{out}(\omega),\\
B^{(1)}_{in}(\omega;\omega_*) &=i(\omega_*-\omega)B_{in}(\omega).
\end{split}
\end{equation}

Using  (\ref{as-2}) and  (\ref{izo}) one can conclude that the transformed  equation (\ref{QNM-sp})  ($A^{(1)}_{in}(\omega;\omega_*)=0$) results in the same QNM spectra (including $\omega=\omega_*$ with solution $1/\phi$). Thus all frequencies of QNM spectra of Regge--Wheeler potential  are frequencies of the QNM spectrum of the new  potential $V^{(1)}$ which
 (and this is exactly what the GJK  have shown) is  the Zerilli potential.

To complete the proof of isospectrality, it remains  to show that no new frequencies appear in the spectrum of the Zerilli potential. Let's assume that there actually exists a solution $\xi=\xi(x;\Omega)$  of the Regge-Wheeler equation
\begin{equation}
\xi''+(\Omega^2-V)\xi=0,
\label{QNM-eq-1}
\end{equation}
such that $\Omega\ne\{\omega_n\}$ (although $\Im ~\Omega<0$) but after (\ref{DT})
$$
\xi^{(1)}(x\to\infty;\Omega;\omega_*)=a^{(1)}_{in}(\Omega;\omega_*){\rm}^{-i\Omega x}+a^{(1)}_{out}(\Omega;\omega_*){\rm}^{i\Omega x},
$$
and $\Omega$ is a solution of the equation $a^{(1)}_{in}(\Omega;\omega_*)=0$.  If such a solution exists, this would mean that DT has generated new frequencies in the QNM spectrum. Fortunately, it does not. For the proof, we use the invertibility of DT: the reverse DT is generated by the function $1/\phi$, i.e.  (\ref{DT}) with  $\sigma^{(1)}=-\sigma$. Therefore
\beqn
\xi=\left(\xi^{(1)}\right)'+\sigma\xi^{(1)},
\enqn
and
\begin{widetext}
\beqn
\xi(x\to+\infty)=-i\left(\Omega+\omega_*\right) a^{(1)}_{in}(\Omega;\omega_*){\rm}^{-i\Omega x}+i\left(\Omega-\omega_*\right) a^{(1)}_{out}(\Omega;\omega_*){\rm}^{i\Omega x},
\enqn
\end{widetext}
i.e. $A_{in}\sim \left(\Omega+\omega_*\right) a^{(1)}_{in}(\Omega;\omega_*)$ therefore the condition $a^{(1)}_{in}(\Omega;\omega_*)=0$ automatically means $A_{in}=0$ therefore $\Omega$ must belong to $\{\omega_n\}$. On a final note, we would like to point out that since $\Im ~\Omega>0$ the condition $\Omega=-\omega_*$ yields no additional roots, concluding our proof of isospectrality.

Now let us discuss the question of a multiple subsequent Darboux transformations.  First of all we note (following the \cite{GJK-2017}) that $\phi$ is the special solution which defines one of the Schwarzschild algebraically special QNM frequencies. This special solution describes  purely ingoing wave, without any reflection  by the black hole wave
potential. The transformed solution  $1/\phi$ describes  purely outgoing wave and, in its turn, is the special solution of the  Zerilli equation. One can use this solution for the next DT, but it would simply revert us back to the initial potential. In order to utilize yet another DT one shall calculate the solution which is linearly independent with  $1/\phi$ solution (the Wronskian of these two solutions must be equal to unity).
After the calculations one will obtain a new solution $\phi^{(1)}$ (note that by using (\ref{DT}) upon $\phi^{(1)}$ will produce nothing but zero, in complete accordance with Property 1):
\begin{equation}
\phi^{(1)}(x;\omega_*)=-\frac{r {\rm e}^{i\omega_*x}}{nr+3M}\int dx \left(n+\frac{3M}{r}\right)^2 {\rm e}^{-2i\omega_*x}.
\label{vtor}
\end{equation}
We calculate this expression for the case $l=n=2$ (if $l=0,\,1$ then $\omega_*=0$) and $\omega_*=-2i/M$). Here is the result:

\begin{widetext}
\begin{equation}
\begin{array}{cc}
\displaystyle{
\phi^{(1)}=\frac{M}{2+3M/r}\left(\frac{(r-2M)^8}{512}\eta_8(r,M){\rm e}^{-i\omega_*x}-1152M^8E_i(-\frac{4r}{M}){\rm e}^{i\omega_*x}\right)},\\
\\
\eta_8(r,M)=-512 r^8+4608 Mr^7-14592M^2r^6+8576M^3r^5+57312M^4r^4-136224M^5r^3+55528M^6r^2+179316M^7r\\
\\-274659M^8, \qquad E'_i(x)={\rm e}^{x}/x.
\end{array}
\label{p1}
\end{equation}
\end{widetext}

The expression (\ref{p1}) is positive for $r>2M$ which makes it suitable for the next DT: $V^{(1)}\to V^{(2)}$ (in the theory of integrable system such transformation is called a {\em binary} DT; we recall that the $V^{(1)}$ is the Zerilli potential $V_{_Z}$).
For that end one must calculate $\sigma^{(1)}=(\log\phi^{(1)})'$ and consider  the behavior of this function at infinity. After the calculations one gets:
$\sigma^{(1)}(x\to\pm\infty)=\mp i\omega_*$, therefore if $\psi^{(1)}$ has the form (\ref{gen-1}) with (\ref{as-2}), (\ref{izo}) then $\psi^{(2)}=(\psi^{(1)})'-\sigma^{(1)}\psi^{(1)}$ has the following asymptotics:
\begin{equation}
\begin{array}{l}
\psi^{(2)}(x\to -\infty;\omega;\omega_*)=B^{(2)}(\omega;\omega_*){\rm e}^{-i\omega x},\\
\\
\psi^{(2)}(x\to +\infty;\omega;\omega_*)=A^{(2)}(\omega;\omega_*){\rm e}^{i\omega x},
\end{array}
\label{p2}
\end{equation}
where
$$
\begin{array}{l}
\displaystyle{
B^{(2)}(\omega;\omega_*)=-i(\omega+\omega_*)B^{(1)}(\omega;\omega_*)},\\
\\
\displaystyle{ A^{(2)}(\omega;\omega_*)=i(\omega+\omega_*)A^{(1)}(\omega;\omega_*).}
\end{array}
$$
Thus, such DT may be  repeated many times. It goes without saying that the new potentials will have quite a
complex form turning the task of discerning its physical interpretation (if any) into a hard and arduous labour. Nevertheless, the isospectral property will be satisfied and the potentials constructed with the aid of a binary DT will all have identical QNM frequencies.

%%%%%%%%%%%%%%1%%%%%%%%%%

Furthermore, using the Darboux transformation, one can construct potentials parametrized by a real parameter $\Lambda\in [0,1]$ $U(x;\omega_*;\Lambda)$ and $W(x;\omega_*;\Lambda)$ with the same QNM spectra and such that:

 (i)  $U(x;\omega_*;\Lambda=0)=V^{(2)}$,   $U(x;\omega_*;\Lambda=1)=V$, where $V$ is the potential of Regge-Wheeler (\ref{R-W});

(ii)     $W(x;\omega_*;\Lambda=1)=V^{(1)}$, where $V^{(1)}$ is the potential of Zerilli.

In order to do this, we shall introduce a new support function $\Phi$:
$$
\Phi=\frac{\Lambda}{\phi}+\left(1-\Lambda\right)\phi^{(1)},
$$
where $\phi$ is given by the (\ref{prop}) and $\phi^{(1)}$ by the (\ref{vtor}). Then we define a new function $\tau=(\log\Phi)'$. It is easy to verify that  $U(x;\omega_*;\Lambda)$ has the form
\begin{equation}
 U=V^{(1)}-2\tau'.
\label{UL}
\end{equation}
Using (\ref{as-2}), (\ref{izo}) and DT; $\psi^{(2)}=(\psi^{(1)})'-\tau\psi^{(1)}$  one can show that new  coefficients $B^{(2)}_{in}$ and $A^{(2)}_{out}$ do not depend on the parameter $\Lambda$ and
$$
\frac{B^{(2)}_{in}}{A^{(2)}_{out}}=\frac{B_{in}}{A_{out}},
$$
where $B_{in}$ and $A_{out}$ are coefficients for the  Regge-Wheeler potential.

The potential $W(x;\omega_*;\Lambda)$  is similarly constructed by the formula
\begin{equation}
W=V-2{\tilde \tau}',
\label{WL}
\end{equation}
with ${\tilde\tau}=(\log{\tilde\Phi})'$ and
$$
{\tilde\Phi}=\phi\left(\Lambda+(1-\Lambda)\int\frac{dx}{\phi^2}\right),
$$
with $\phi$ from the (\ref{prop}). We'd like to stress that both $U$ and $W$ from (\ref{UL}), (\ref{WL}) have the same QNM spectra as Regge-Wheeler ($V$)  and Zerilli ($V^{(1)}$) potentials.

In fact, DT can be implemented without the use of special solutions of the type (\ref{prop}). Let $\omega_n$ be QNM spectra, which by definition is a solution of the equation $A_{in}(\omega_n)=0$.
In other words, the eigenfunctions $\psi_n(x;\omega_n)$ have asymptotic behavior:
$$
\begin{array}{l}
\displaystyle{
\psi_n(x\to +\infty)=A_{out}(\omega_n){\rm e}^{i\omega_n x}},\\
\\
\displaystyle{
 \psi_n(x\to -\infty)=B_{in}(\omega_n){\rm e}^{-i\omega_n x}.}
\end{array}
$$
If for some number $n$ (we put it $n=1$) the solution $\psi_1(x;\omega_1)>0$ for $-\infty<x<+\infty$ then one can use $\psi_1$ as the support function for the DT. So for the $n\ne 1$ one gets
$$
\psi^{(1)}_n(x;\omega_n;\omega_1)=\frac{\psi'_n(x;\omega_n)\psi_1(x;\omega_1)-\psi_n(x;\omega_n)\psi'_1(x;\omega_1)}{\psi_1(x;\omega_1)},
$$
and one can verify that
$$
\begin{array}{l}
\psi^{(1)}_n(x\to +\infty;\omega_n;\omega_1)=A^{(1)}_{out}(\omega_n;\omega_1){\rm e}^{i\omega_n x},\\
\\
\psi^{(1)}_n(x\to -\infty;\omega_n;\omega_1)=B^{(1)}_{in}(\omega_n;\omega_1){\rm e}^{-i\omega_n x},
\end{array}
$$
with
$$
\begin{array}{l}
A^{(1)}_{out}(\omega_n;\omega_1)=i(\omega_n-\omega_1)A_{out}(\omega_n),\\
\\
B^{(1)}_{in}(\omega_n;\omega_1)=-i(\omega_n-\omega_1)B_{in}(\omega_n).
\end{array}
$$
If $\psi^{(1)}_2(x;\omega_2;\omega_1)$ is  positive for all $x\in \mathbb{R}$ , then it can be used as a support function for the next Darboux transformation and as a way to explicitly constructing the Crum formulas.

\section{Generalized Darboux Transformation} \label{sec:GDT}

Let us now ponder (\ref{GDT}). Before we begin, let us emphasize one important point: we do {\em not} claim that GDT can under no circumstance be used to connect the isospectral potentials, for such a claim would not only be exceedingly strange, but patently false as well. In fact, the GDT \eqref{GDT} naturally arises when one attempts to generate long iterative chains of ordinary DTs \cite{Shah_Whiting}. What we are arguing, though, is that the resulting relationship can be called a DT only {\em formally}, due to coefficients $A$ and $B$ in \eqref{GDT} becoming explicitly dependent on the spectral parameter $\lambda$ of the {\em dressed} function. To put it in other words, different $\psi(x;\lambda)$ necessitate different transformation rules. For example, consider a sample collection of multiple solutions $\psi=\psi(x;\l_i)$ of \eqref{Scr} with a given potential $V(x)$ but different eigenvalues $\l_i$, and let $\l \neq \l_i$ for $\forall i$. Then, conduct two consecutive over $\psi(x;\l)$, using two support functions from our collection, say, $\psi_1$ and $\psi_2$. A new, doubly dressed function $\psi^{(2)} = \psi^{(2)}(x; \l; \l_1; \l_2)$ can be calculated via the Crum's formulas \cite{Crum}, but it can also be derived from the original $\psi$ with the aid of the first order linear differential operator:
\beq \label{tilde_psi}
\psi \to \psi^{(2)} \equiv \tilde \psi = A \psi'-B \psi,
\enq
where
\beq \label{A_and_B}
A = \frac{\left(\l_2-\l_1\right) \psi_1 \psi_2}{W\left(\psi_2,\psi_1\right)}, \quad B = \l + \frac{\l_2 \psi_2 \psi'_1 - \l_1 \psi_1 \psi'_2}{W(\psi_2,\psi_1)},
\enq
and $W(\psi_2,\psi_2)$ is the Wronskian of those two solutions. From \eqref{A_and_B} it is immediately apparent that $B$ explicitly depends on $\l$. Similarly, a chain of three subsequent DTs with support functions $\psi_1$, $\psi_2$ and $\psi_3$ is going to produce $\psi^{(3)} \equiv \tilde \psi$, which would again satisfy \eqref{tilde_psi}, but this time with $A$ gaining its own dependence upon $\l$:
\beq \label{A_3}
A(x;\l;\l_1;\l_2;\l_3) = V(x) -\l + \frac{W_1\left(\psi_3,\psi_2,\psi_1\right)}{W\left(\psi_3,\psi_2,\psi_1\right)},
\enq
where $W_1$ is the following specially modified Wronskian:
\beqn
W_1\left(\psi_3,\psi_2,\psi_1\right)=\left|
\begin{array}{ccc}
\psi'''_3 & \psi'''_2 & \psi'''_1\\
\psi''_3 & \psi''_2 & \psi''_1\\
\psi_3 & \psi_2 & \psi_1
\end{array} \right|.
\enqn
The explicit form of coefficient $B=B(x;\l;\l_1;\l_2;\l_3)$ can be calculated in a similar fashion, but will be omitted here due to its overall cumbersomeness.

The aforementioned process can be repeated as many times as we have functions in our collection, but should already be apparent that the ``GDT'' \eqref{GDT} (or \eqref{tilde_psi}) indeed emerges when considering the chains of ordinary DTs, and, naturally, it must preserve the required isospectrality (since the individual DTs do). However, there is a caveat: the coefficients $A$ and $B$ in GDT {\em have} to depend upon the spectral parameter of a dressed function. In general, we shall expect that the very presence of such a dependency would make the transformation \eqref{tilde_psi} formal and not very useful. This might sound prohibitively pessimistic, especially in view of the DT chains counterexample we have just discussed, but it is important to point out how exceptionally special, if not downright unique, this example is. Of course, there might still exist some other cases of GDTs with $\l$-dependent coefficients, which are not reducible to a mere chain of ordinary DTs. If they do, this would be a very remarkable and interesting fact, deserving of a comprehensive study of its own! Nevertheless, below we will consider a more traditional GDT with coefficients $A=A(x;\mu)$ and $B=B(x;\mu)$, which depend on the spectral parameter $\mu$, but of a support function, and support function only.

So, let us carry on with such a GDT. In order to satisfy Property 1, it is instructive to rewrite it in the form:
\begin{equation}
 \psi^{(1)}=A(x)\left(\psi'-\sigma\psi\right).
 \label{NGDT}
 \end{equation}
Substituting into the transformed equation (this time the new potentials are denoted with tildes, lest we confuse them with the potentials in formulas (\ref{DT})):
$$
\left(\psi^{(1)}\right)''+\left(\lambda-{\tilde V}\right)\psi^{(1)}=0,
$$
after some calculations we get:
\begin{equation}
M(x;\mu)\psi'(x;\lambda) + N(x;\mu;\lambda) \psi(x;\lambda) =0.
\label{soot}
\end{equation}

We are assuming that the potential $V=V(x)$ is not a constant; this implies that the functions $\psi$ and $\psi'$ are linearly independent functions. Hence, the only way \eqref{soot} can be satisfied is by setting:
$$
M(x;\mu)=0, \qquad N(x;\mu;\lambda)=0.
$$
From the first relation we find
\begin{equation}
 {\tilde V(x;\mu)}=V-2\sigma'+\frac{A''-2\sigma A}{A},
 \label{tilV}
 \end{equation}
where we used the  (\ref{Ric}). Substituting  (\ref{tilV}) into the  (\ref{soot}) we have
$$
2A'(x)(\mu-\lambda)\psi=0,
$$
from which it follows that either  (i) $A={\rm const}$ or  (ii) $\lambda=\mu$. The former is nothing else but a standard (not generalized)  DT, whereas in the latter case we are restricting our attention solely to zero modes and, therefore, are incapable of gathering any information about the rest of the QNM spectrum.
\\
{\bf Comment.} The relation (\ref{tilV}) was certainly seen by the authors of \cite{GJK-2017}, but they interpreted it in a rather ingenious way: they used (\ref{NGDT}), but   as a $\sigma$ they used expression  $\sigma = u '/ u $ where $ u $  is the solution of the equation not with the potential $ V $, but with the potential $ V + c / A^2 $, where $ c $ is an arbitrary constant. Unfortunately, the introduction of such an auxiliary problem does not change anything: the function $ u $ must correspond to the same value of the spectral parameter $\lambda $. In other words, GJK  (and us)  are forced to choose the case (ii). We note that if $ A $ = const and we denote $ c / A^2- \lambda = - \mu $, then, of course, we arrive at the usual DT, simultaneously acquiring the isospectral property ($ c $ is arbitrary, which means that $ \mu $ is arbitrary too).

In \cite{GJK-2017} generalized Darboux transformation  was used to establish the connection between  short-ranged and long-ranged potentials,    with a subsequent generalization to perturbation theory for Kerr black holes.  In fact GJK  used the arbitrariness of the function  $A(x)$.   We have already established that this only works for $\lambda = \mu $.
 In fact, under this condition, GDT (\ref {NGDT}) allows one  to connect   {\em any two ad hoc potentials}. Let us show this.

We have already seen that if  $\lambda=\mu$ and $\psi=\phi$ then  (\ref{NGDT}) result in zero: $\phi^{(1)}=0$.    In the previous section, we have shown a way to rectify this hindrance: for that end let ${\hat\phi}(x;\mu)$ be a solution,  linearly independent from $\phi(x;\mu)$ (so that the Wronskian $W({\hat\phi},\phi)=1$):  ${\hat\phi}=\phi\int dx\phi^{-2}$. Substituting this new function into (\ref{NGDT}) yields ${\hat\phi}^{(1)}=A(x)/\phi$. The function ${\hat\phi}^{(1)}$ has to satisfy the equation
\begin{equation}
\left({\hat\phi}^{(1)}\right)''+(\mu-W(x)){\hat\phi}^{(1)}=0,
\label{111}
\end{equation}
where $W(x)$ is {\bf arbitrary} ad hoc potential. It is easy to check that if $A'=0$ then $W(x)=V^{(1)}(x;\mu)$. In the case of general position one have the equation for unknown function $A(x)$:
\begin{equation}
A''-2\sigma A'+\left(V^{(1)}-W\right)A=0,
\label{122}
\end{equation}
where $V^{(1)}$ is defined by (\ref{DT}), thus rendering the reason why the condition $A={\rm const}$ results in $W(x)=V^{(1)}(x;\mu)$ quite  clear.  In addition, we also note that it is  possible to ``invert'' our procedure by defining $W(x)=V^{(1)}(x;\mu)$; in this case the equation  (\ref{122}) will end up being trivially integrable with respect to  $A(x)$. By dividing this expression by $\phi $, we obtain
$\phi^{-1}\int dx \phi^2$ -- the solution which is linearly independent with $1/\phi$.   In the general case of an arbitrary $W(x)$, the equation (\ref{122}) can not be easily integrated, but it is of small consequence, since all we really need to know is that the solution of the equation (\ref{122}) does exist, and that for any non-singular $W$ and $V^{(1)}$ and positively defined $\phi (x; \mu)$ this solution will be defined everywhere.
So, for $\lambda = \mu $ GDT (\ref{NGDT}) indeed allows us to link  $V(x)$   with an arbitrary  potential $W(x)$. The only snag is that this link is completely useless from a practical point of view,  because the equation (\ref{122}) will be explicitly integrable if and only if we can explicitly integrate the equation (\ref{111}), and if this is the case, then {\em why bother with GDT at all?} In particular, it is always possible to formally associate the potentials from the Bardeen-Press and the Regge-Wheeler equations or, for that matter, with any other ``nice'' potential (such as the short ranged potential from the Sasaki-Nakamura equation \cite{Sasaki}) that strikes our fancy. But what would be the point?.. The presence of these ``Darboux-like'' connections is just not well-suited for the task of finding the exact solutions or for determining the QNM spectrum. Now, we would like to stress yet again that we are not claiming that GDT is useless per se nor that it is of no interest in general. What we criticise here is the alleged effectiveness of GDT \eqref{tilde_psi} in the framework of specific mathematical problem: calculating the unknown QNM spectrum for a potential, using the known QNM spectrum of yet another potential.

One seemingly obvious way to try to make (\ref{NGDT}) more effective would be to dispense with the Property 1 altogether. The problem is, what we would end up with will be something that can hardly be called a Darboux transformations anymore. And this problem has nothing to do with the semantics, too: all the seemingly miraculous powers of DT hang upon just one reason: {\em its isospectrality}. And it is exactly this crucial property that will be utterly destroyed by abandoning the Property 1 (see, however, the discussion at the end of Sec. \ref{sec:Frequency}).

Finally, we can try to bypass the property $\lambda = \mu $ in the following way: we define the function $\Psi (x; \mu; \lambda) $ by the relation
\begin{equation}
\Psi=A(x){\bf q}\psi+B(x){\bf q}\left(\psi\int\frac{dx}{\psi^2}\right),
\label{Psi}
\end{equation}
where  ${\bf q}=d/dx-\sigma(x;\mu)$, $\psi=\psi(x;\lambda)$, $A(x)$ and   $B(x)$ are two as yet undefined functions. We require that
\begin{equation}
\Psi''+(\lambda-W(x;\mu))\Psi=0.
\label{eqPsi}
\end{equation}
For convenience we define the operators $L_m(A,B)$ such that
\begin{equation}
L_m(A,B)=\frac{d^m A}{dx^m}+g \frac{d^m B}{dx^m},
\label{Lm}
\end{equation}
with $g'=-1/\psi^2$. Substituting (\ref{Psi}) into the (\ref{eqPsi}) we get
\begin{equation}
H_1\psi'-H_0\psi+\frac{H_{-1}}{\psi}=0,
\label{eqq}
\end{equation}
where
$$
\begin{array}{c}
H_1=L_2(A,B)-2\sigma L_1(A,B) +\left(V^{(1)}-W\right) L_0(A,B),\\
\\
H_0=\sigma L_2(A,B)-\left(V^{(1)}+V-2\lambda\right) L_1(A,B)+\\
\\
+\sigma\left(V^{(1)}-W\right) L_0(A,B),\\
\\
H_{_{-1}}=B''-2B'\sigma+\left(V^{(1)}-W\right)B.
\end{array}
$$
Now we end up with three equations to solve:
$$
H_1=H_0=H_{-1}=0.
$$

The rather cumbersome calculations results in the condition
 $$
 2\left(\lambda-\mu\right)\left(B'g+A'\right)=0.
 $$
We explicitly want to avoid the case $\lambda=\mu$. Therefore:
$$
\begin{array}{l}
\displaystyle{
\left(\log B'\right)'=2\sigma'-\frac{B}{\psi^2 L_0(A,B)},}\\
\\
\displaystyle{ A(x)=-\int dx B'(x)\int^x\frac{dy}{\psi^2(y)},}
\end{array}
$$
and the sought after expression for the potential takes the following form:
\begin{equation}
W=V-2\sigma-\frac{B'}{\psi^2 L_0(A,B)}.
\label{WW}
\end{equation}
Since both $\psi$ and $B$ depend on  $\lambda$, we conclude that so would the potential $W=W(x;\mu;\lambda)$. This violates the Property 2, and so the transformation (\ref{Psi}), (\ref{WW}) ends up being just some linear substitution instead of a fully-fledged Darboux transformation proper.

\section{Frequency-dependent potentials} \label{sec:Frequency}

At last, let us entertain a possibility of isospectral DT existing for those potentials that are frequency-dependent (e.g., the Bardeen-Press potential, the Chandrasekhar-Detweiler potential for the Kerr black holes \cite{CD, Detweiler, Nakamura}, or the potentials for the Teukolsky and Sasaki-Nakamura equations). Naturally, the study of such potentials commands a lot of attention and interest not just from the physical but from the mathematical point of view as well.

We will consider two potentials $V(x;\lambda)$, $V(x;\mu)$. Let $\psi(x;\lambda)$ and $\phi(x;\mu)$ be two solutions of the following differential equations:
\begin{equation}
\begin{array}{l}
\psi''(x;\lambda)+\left(\lambda-V(x;\lambda)\right)\psi(x;\lambda)=0,\\
\\
 \phi''(x;\mu)+\left(\mu-V(x;\mu)\right)\phi(x;\mu)=0.
\end{array}
\label{mu-lam}
\end{equation}
As usual, $\sigma(x;\mu)=(\log\phi(x;\mu))'$. Then, the Darboux-like transformation has the form:
\begin{widetext}
\begin{equation}
\begin{array}{l}
\displaystyle{
\psi^{(1)}(x;\lambda;\mu)=\beta(x;\lambda;\mu)\left(\psi'(x;\lambda)-\sigma(x;\mu)\psi(x;\lambda)\right),}\\
\\
\displaystyle{V^{(1)}(x;\lambda;\mu)=V(x;\lambda)-2\sigma'(x;\mu)+\frac{\beta''(x;\lambda;\mu)-2\sigma(x;\mu)\beta'(x;\lambda;\mu)}{\beta(x;\lambda;\mu)},}\\
\\
\displaystyle{
\beta(x;\lambda;\mu)=\left(V(x;\lambda)-V(x;\mu)+\mu-\lambda\right)^{-1/2}.}
\end{array}
\label{DT-mu-lam}
\end{equation}
\end{widetext}
Well, but what about the isospectrality? For the frequency-dependent potentials, the Property 2 is violated. This circumstance creates the hope that an effective generalized isospectral Darboux transformation is still possible. Unfortunately, this hope turns out to be illusory.

Lets consider a specific example of the application of formulas (\ref{DT-mu-lam})to demonstrate why these transformations can not be isospectral.

Consider two one-soliton potentials with $\lambda=-k^2$ and $\mu=-\kappa^2$:
 \begin{equation}
 V(x;\lambda)=-\frac{2k^2}{\cosh^2 kx},\qquad V(x;\mu)=-\frac{2\kappa^2}{\cosh^2 \kappa x}.
 \label{soliton}
 \end{equation}
 In ordinary radial variables the potential $V(x;\lambda)$ has the form:
 $$
 V=-\frac{8k^2(r-2M)^{4Mk}}{(r-2M)^{4Mk}\left[2+(r-2M)^{4Mk}{\rm e}^{2kr}\right]+{\rm e}^{-2kr}},
 $$
so
$$
V(r\to 2M)=0,
$$
and
$$
 V(r\to\infty)\to -8k^2{\rm e}^{-2kr}r^{-4Mk}\to 0.
$$
Such potentials have no apparent physical meaning in the theory of black holes; still, they are quite useful as a sort of a mathematical test subject.

The  solution in the form (\ref{9-10}) of the equation  (\ref{Scr}) with the potential $V(x;\lambda)$ (\ref{soliton}) and the frequency  $\omega$ has the form
\begin{equation}
\begin{array}{l}
\displaystyle{
\psi_+(x;\omega)=c_1\left(i\omega+k\tanh kx\right){\rm e}^{-i\omega x}+}\\
\\
\displaystyle{
+c_2\left(k\tanh kx-i\omega\right){\rm e}^{i\omega x},\qquad x>0},\\
\\
\displaystyle{
\psi_-(x;\omega)=c_3\left(i\omega+k\tanh kx\right){\rm e}^{-i\omega x},\qquad x<0.}
\end{array}
\label{solit-l}
\end{equation}
Using (\ref{10-9}) one gets
$$
\begin{array}{l}
A_{in}(\omega;k)=c_1\left(i\omega+|k|\right),\qquad A_{out}(\omega)=c_2\left(|k|-i\omega\right),\\
\\
B_{in}(\omega;k)=c_3\left(i\omega-|k|\right).
\end{array}
$$
 The equation (\ref{QNM-sp}) results in $\omega_1=i|k|$, therefore we do not have a QNM spectra for the potential (\ref{soliton}) (in what follows, we are going to assume without a loss of generality that $k>0$). The continuity condition results in
 $c_3=c_1-c_2$ and   $\omega^2+k^2=0$ that is valid for the $\omega=\omega_1$.

Let $k>\kappa>0$.  The $\beta^{-2}(x)$ is an even function having a minimum at  x = 0 ($\beta^{-2}(0)=\kappa^2-k^2<0$ ) and $\beta^{-2}(|x|\to\infty)=k^2-\kappa^2>0$. In addition, the equation $\beta^{-2}(x)=0$ has two symmetric roots: $|x|=x_*$ such that  $\beta^{-2}(x)>0$ if $|x|>x_*$ and $\beta^{-2}(x)<0$ if $|x|<x_*$.
Thus the potential  $V^{(1)}$ (\ref{DT-mu-lam}) is real-valued when $|x|>x_*$, complex-valued for  $|x|<x_*$ and singular at $|x|=x_*$ (a behaviour that is not dissimilar to the ones of the Chandrasekhar and  Detweiler Kerr wave equations with short range potentials, singular at certain $r$).  Although the model we just constructed does not appear to be physical, it is nevertheless a good primer to use for verifying the isospectrality (or the lack thereof) of the transformations (\ref{DT-mu-lam}).

For that end we choose the exact solution of the equation with $V(x;\mu)$ as $\phi=1/\cosh(\kappa x)$ so  $\sigma=-\kappa\tanh\,\kappa x$. Substituting  $\sigma$ and (\ref{solit-l}) into the  (\ref{DT-mu-lam}) and calculating the limit with $x\to+\infty$  we get
 $$
 A^{(1)}_{in}(\omega;k;\kappa)=c_1\left(i\omega+k\right)\left(\kappa-i\omega\right),
 $$
 and  the equation (\ref{QNM-sp}) results in $\omega_1=-i\kappa$ so we have one new QNM frequency. Thus we conclude that whatever the transformation (\ref{DT-mu-lam}) is, isospectral it is not.

Granted, it would be a mistake to consider all non-isospectral ``DT'' as completely useless and unapplicable. A beautiful rebuttal to such a view can be gleamed from the case of spherical functions $\theta_{_{lm}}$, the solutions of the following equation:
\begin{equation}
\displaystyle{
\frac{1}{\sin x}\frac{d}{dx}\left(\sin x\frac{d\theta_{_{lm}}}{dx}\right)-\frac{m^2}{\sin^2 x}~\theta_{_{lm}}+l(l+1)\theta_{_{lm}}=0.
}
\label{NI1}
\end{equation}

The spherical functions are directly proportional to the associated Legendre polynomials $P_l^m(\cos x)$, where $x$ is the azimuth angle, $l$ and $m$ are, subsequently, the orbital and the magnetic numbers. Introducing the substitution $\theta_{_{lm}} =\left(\sin x\right)^{-1/2}\Psi_{_{lm}}$, we reduce \eqref{NI1} to the following Schr\"odinger equation:
\begin{equation}
\displaystyle{-\frac{d^2\Psi_{_{lm}}}{dx^2}+\frac{m^2-\frac{1}{4}}{\sin^2 x}~\Psi_{_{lm}}=\left(\frac{1}{4}+l(l+1)\right)\Psi_{_{lm}}.
}
\label{NI2}
\end{equation}

In this equation we encounter a new shape-invariant potential $\sin^{-2} x$, which is preserved by DT, albeit with a different multiple. This implies that DT for this equation alters the magnetic number $m$, which is done by two possible routes: $m\to m\pm 1$. In other words, DT here literally goes through all the wave functions for a fixed orbital number $l$ (i.e. through the entire subspace of irreducible representation of three-dimensional rotations with weight $l$).

Another good example is a three-dimensional potential $V(r)$, which possesses a central symmetry. Here is the equation for the radially symmetric wave function $R(r)$, with a given orbital quantum number $l$:
\begin{equation}
\displaystyle{
\frac{1}{r^2}\frac{d}{dr}\left(r^2\frac{dR}{dr}\right)+\left(\lambda-V(r)-\frac{l(l+1)}{r^2}\right)R=0,
}
\label{NI}
\end{equation}
and if we introduce a ``dilatory'' variable $\xi=\ln r$, we can reduce \eqref{NI} to yet another Schr\"odinger equation, defined for all $\xi \in \R$:
\begin{equation}
\displaystyle{
\frac{d^2\Psi}{d\xi^2}=\left(U(\xi)-\epsilon\right)\Psi,
}
\label{NI3}
\end{equation}
where
\beqn
R={\rm e}^{-\xi/2}\Psi(\xi),\,\, U(\xi)={\rm e}^{2\xi}\left(V(\xi)-\lambda\right),\,\, \epsilon=-\frac{1}{4}-l(l+1).
\enqn

For \eqref{NI3} one can yet again construct ``true'', isospectrality-preserving DT, but this time the role of the spectral parameter will be played by $\epsilon$, associated with the orbital quantum number $l$. Thus, in this particular case DT works by establishing the relationship between different wave functions that correspond to {\em different} irreducible representations and {\em different} potentials (which are also related via DT). Furthermore, if those different potentials are additionally tied to each other by some simple algebraic relationship -- if, for example, they end up being the same shape-invariant potential, -- then the Schr\"odinger equation becomes completely integrable.

These two examples are but a small sample of interesting problems where the non-isospectral Darboux-like transformations lend a much welcome aid (for example, for the two-dimensional problems we might name the Moutard \cite{Moutard} and Laplace \cite{Laplace} transformations, who are exceptionally useful in the theory of (1+2) integrable hierarchies, despite having no isospectral symmetry whatsoever). However, we would like to remind our reader that it is the QNM for the black holes that was the primary focus of this article, and it is for this particular problem that the Darboux transformation proves to be a powerful tool in the arsenal of the physicist -- provided, that this DT is additionally equipped with the property of isospectrality. Which explains why we are so interested in this property to begin with!

\section{Conclusion}

In this paper we have studied in detail DT, BDT, GDT plus the DT for the frequency-dependent potentials in black hole perturbation theory. Our findings can be summarized as follows:
\newline
(i) The isospectrality of the Darboux transformations is completely proven.
\newline
(ii) The isospectrality of the binary Darboux transformations is completely proven, in stark contrast to the standard quantum mechanics where (with some rare exceptions, mentioned above) the binary DT do not possess this property.
\newline
(iii) The generalized DT has not this property and are valid only for solutions with the fixed frequency but in this case one can connect any ad hoc potentials via GDT.
\newline
(iv) We have constructed an analog of the Darboux transformations for frequency-dependent potentials and have shown that such transformations are not isospectral, so their usefulness for QNM  spectral computation problems appears to be rather doubtful.

In conclusion, we can only agree with the following opinion of the authors of  the paper \cite{GJK-2017}:  ``It may even be that future efforts should look beyond the
Darboux method. Other methods of seeking isospectral potentials, such as the one described in Abraham  and  Moses \cite{Abraham_Moses}, exist''. It must be owned that, as the authors of this article, and as mathematicians who has been working with DT and its various applications for quite a long time, we would have loved nothing more then to arrive to something more optimistic then the conclusion presented here. We sincerely hope that the future studies will provide a more positive take on the effectiveness of GDT with $\l$-dependent  coefficients (see the discussion at the beginning of Sec. \ref{sec:GDT}) for the fascinating problem of QNM spectrum's derivation. But at this juncture it would probably be prudent not to count on that.

%%%%%%%%%%%%%%%%%%%%%%%%%%%%%%%%%%%%%%%%%%%%%%%%%%%%%%%%%

\begin{acknowledgments}
The work was supported from the Russian Academic Excellence Project at the Immanuel Kant Baltic Federal University, and by the project 1.4539.2017/8.9 (MES, Russia). The authors would also like to express their gratitude to anonymous referee for substantial critique of the original version of this article, which led to a significant improvement of the manuscript.
\end{acknowledgments}
$$
{}
$$

%\newpage %Just because of unusual number of tables stacked at end
\bibliography{apssamp}% Produces the bibliography via BibTeX.

\end{document}